\pdfoutput=1

\documentclass[sigconf,natbib=true,anonymous=false]{acmart}

\setcopyright{none}
\renewcommand\footnotetextcopyrightpermission[1]{}

\usepackage[T1]{fontenc}

\usepackage[hang,flushmargin]{footmisc}

\usepackage{url}            
\usepackage{booktabs}       
\usepackage{amsfonts}       
\usepackage{nicefrac}       
\usepackage{microtype}      
\usepackage{amsmath}
\usepackage{enumitem}
\usepackage{graphicx}       
\usepackage{caption}
\usepackage{subcaption}
\usepackage{threeparttable}
\usepackage{color, colortbl}
\usepackage{threeparttable,booktabs}
\usepackage{multirow}
\usepackage{balance}
\usepackage{xspace}
\usepackage{balance}

\usepackage[frozencache=true,cachedir=minted-cache]{minted}
\definecolor{LightGray}{gray}{0.9}
\usepackage{ragged2e}

\newcommand{\umb}{\mbox{UMBRELA}\xspace}

\newcommand\ignore[1]{}

\title{A Large-Scale Study of Relevance Assessments with\\ Large Language Models:\ An Initial Look}

\author{Shivani Upadhyay}
\affiliation{%
  \institution{University of Waterloo}
  \state{Ontario}
  \country{Canada}
}

\author{Ronak Pradeep}
\affiliation{%
  \institution{University of Waterloo}
  \state{Ontario}
  \country{Canada}
}

\author{Nandan Thakur}
\affiliation{%
  \institution{University of Waterloo}
  \state{Ontario}
  \country{Canada}
}

\author{Daniel Campos}
\affiliation{%
  \institution{Snowflake}
  \city{San Mateo}
  \country{USA}
}

\author{Nick Craswell}
\affiliation{%
  \institution{Microsoft}
  \city{Seattle}
  \country{USA}
}

\author{Ian Soboroff}
\affiliation{%
  \institution{NIST}
  \city{Gaithersburg}
  \country{USA}
}

\author{Hoa Trang Dang}
\affiliation{%
  \institution{NIST}
  \city{Gaithersburg}
  \country{USA}
}

\author{Jimmy Lin}
\affiliation{%
  \institution{University of Waterloo}
  \state{Ontario}
  \country{Canada}
}

\begin{document}

\begin{abstract}
The application of large language models to provide relevance assessments presents exciting opportunities to advance information retrieval, natural language processing, and beyond, but to date many unknowns remain.
This paper reports on the results of a large-scale evaluation (the TREC 2024 RAG Track) where four different relevance assessment approaches were deployed {\it in situ}:\ the ``standard'' fully manual process that NIST has implemented for decades and three different alternatives that take advantage of LLMs to different extents using the open-source \umb tool.
This setup allows us to correlate system rankings induced by the different approaches to characterize tradeoffs between cost and quality.
We find that in terms of nDCG@20, nDCG@100, and Recall@100, system rankings induced by automatically generated relevance assessments from \umb correlate highly with those induced by fully manual assessments across a diverse set of 77 runs from 19 teams.
Our results suggest that automatically generated \umb judgments can replace fully manual judgments to accurately capture run-level effectiveness. 
Surprisingly, we find that LLM assistance does {\it not} appear to increase correlation with fully manual assessments, suggesting that costs associated with human-in-the-loop processes do not bring obvious tangible benefits.
Overall, human assessors appear to be stricter than \umb in applying relevance criteria.
Our work validates the use of LLMs in academic TREC-style evaluations and provides the foundation for future studies.
\end{abstract}

\renewcommand{\shortauthors}{}
\pagestyle{empty}

\maketitle

\section{Introduction}

Relevance assessments are critical for evaluating information access systems, providing guidance for model training in information retrieval (IR), natural language processing (NLP), and beyond.
Acquiring relevance assessments from humans, of course, is an expensive proposition, not only in terms of compensation for the assessors, but also overhead in articulating clear guidelines, coordinating large-scale annotation efforts, ensuring consistent quality, etc.
The advent of large language models (LLMs) provides an opportunity to automate these assessments, potentially accelerating improvements in retrieval methods~\cite{Faggioli_etal_ICTIR2023}---although see~\citet{Soboroff:2409.15133:2024} for a contrary position.

Most recently, \citet{ThomasPaul_etal_SIGIR2024} revealed that LLMs have been used to provide relevance assessments at Bing since late 2022.
While the authors quite clearly assert that LLMs can accurately predict searcher preferences in the context of Bing, they began from the industry perspective and retrospectively ``backported'' their proposed fully automatic technique to an old TREC collection for validation.
In this work, we adopt a complementary strategy, exploring different applications of LLMs to generate relevance assessments {\it in situ}, directly deployed in a large-scale retrieval-augmented generation (RAG) evaluation organized by the National Institute of Standards and Technology (NIST) as part of the Text Retrieval Conference (TREC) series.
Now in its 33rd year, TREC is widely acknowledged as representing the ``gold standard'' in academic IR evaluations.

We examined three different evaluation approaches that vary in levels of LLM involvement, which are compared against fully manual relevance assessments.
Our study used the \umb tool~\cite{Upadhyay_etal_arXiv2024}, which has been previously validated as a successful reproduction of \citet{ThomasPaul_etal_SIGIR2024}.
Given this setup, we explored the following research questions:

\begin{itemize}

\item[{\bf RQ1}] To what extent can automatically generated relevance assessments from LLMs replace NIST assessors?
Specifically, we examined three scenarios involving different levels of LLM involvement:\ fully automatic, manual post-editing, and manual filtering.

\item[{\bf RQ2}] Different levels of LLM involvement lead to different tradeoffs between cost and quality in relevance assessments.
Can we characterize these tradeoffs?

\item[{\bf RQ3}] Are there any quantitative or qualitative differences between human versus LLM judgments?
Are these differences affected by different levels of LLM involvement?

\end{itemize}

\noindent In this work, we analyzed a diverse set of 77 runs from 19 teams that contributed to the TREC 2024 RAG Track.
Our findings can be summarized as follows:

\begin{itemize}[leftmargin=*]

\item For RQ1, we find that system rankings induced by automatically generated relevance assessments using \umb correlate highly with those induced by manual assessments in terms of nDCG@20, nDCG@100, and Recall@100.

\item For RQ2, we find that, surprisingly, LLM assistance does {\it not} appear to increase correlation with fully manual assessments.
Thus, the additional costs associated with hybrid human--LLM solutions do not appear to have obvious tangible benefits.

\item For RQ3, analyses suggest that assessors apply stricter relevance criteria than \umb.
We find cases where the LLM draws inferences that humans would consider unwarranted, and vice versa, but overall, the first case is more common.

\end{itemize}

\noindent The contribution of this work is, to our knowledge, the first large-scale study of different {\it in situ} approaches to using automatically generated LLM assessments, where system rankings induced by these approaches are correlated against reference rankings generated by fully manual assessments.
This represents a first step outside the industry setting to validate methods for automatically generating relevance assessments using LLMs.
In this context, our work contributes to a long tradition of meta-evaluation in the information retrieval literature, evaluations that evaluate evaluation methodologies.

\section{Background and Related Work}
\label{section:background}

Relevance assessments, alternatively called labels, judgments, annotations, or qrels (and used interchangeably throughout this paper), drive progress in information retrieval and beyond by enabling the training of better models and the evaluation of their quality.
These assessments can come from a variety of sources.
Explicit high-quality labels can be provided by humans, for example, editors hired by web search engine companies or retired intelligence analysts in the case of TREC evaluations~\cite{Harman_2011}.
Alternatively, implicit and often noisy labels can be extracted from user behavior logs~\cite{Joachims05}, and most recently, the field has seen the growing popularity of synthetically generated labels~\cite{Bonifacio_etal_SIGIR2022,Dai_etal_ICLR2023}.
Despite these diverse approaches, human-generated labels are widely acknowledged to represent the best source of data, against which all other techniques are compared.
In the academic context, relevance assessments organized by NIST for TREC are recognized as the gold standard.

\citet{Faggioli_etal_ICTIR2023} (whose explorations began in early 2023, shortly after the release of ChatGPT) were among the earliest to discuss prompting LLMs to provide relevance judgments.
Their paper explored a spectrum of human--machine collaboration strategies, articulating arguments for and against different approaches.
We can readily situate the various conditions in our study along their proposed spectrum.

Far more than academic speculation, \citet{ThomasPaul_etal_SIGIR2024} quite clearly stated, in the context of the Bing search engine:\ ``We have been using LLMs, in conjunction with expert human labellers, for most of our offline metrics since late 2022.''
They assert that LLMs can accurately predict searcher preferences and that LLMs are as accurate as human labellers for evaluating systems.
These claims were validated using an old TREC collection (2004 Robust Track), where an ``in-house'' version of GPT-4 was used (re-)label old queries and documents with different prompt configurations.

While \citet{ThomasPaul_etal_SIGIR2024} was no doubt enlightening in revealing industry practices, we can point to a few shortcomings.
First, the retrospective experimental design is not ideal---the authors were essentially ``backporting'' a modern technique to a collection that is 20 years old.
Second, there is a concern about data contamination~\cite{Deng:2311.09783:2024}, whereby publicly available relevance labels might have (inadvertently) become incorporated into LLM pretraining data.\footnote{To us, the first and second issues are related but orthogonal:\ the first concern is about the methodology itself, while the second is about the data used.}
Third, while the authors described processes at Bing that involve human intervention, their published experiments were fully automatic, and hence do not probe the potential synergies of human--machine hybrids.
Finally, \citet{ThomasPaul_etal_SIGIR2024} reported using an ``in-house'' custom version of GPT-4 that is not publicly available.

The last limitation was addressed by \citet{Upadhyay_etal_arXiv2024}, who built and released \umb, which is a recursive acronym that stands for {UMbrela} is the {Bing} {RELevance} {Assessor}.
As the name suggests, \umb is an open-source toolkit that successfully reproduced the techniques described in \citet{ThomasPaul_etal_SIGIR2024} and affirmed the paper's main claims.
While the tool still depends on a proprietary large language model, at least the model is publicly accessible.
\umb was used by two teams independently in submissions to the LLM4Eval Workshop at SIGIR 2024~\cite{Rahmani:2408.08896:2024}, which provides confidence that internal work inside Bing can be reproduced by the broader community.

This work specifically addresses all the remaining identified shortcomings of \citet{ThomasPaul_etal_SIGIR2024}:\
We report on an {\it in situ} evaluation where four different approaches to relevance assessments were deployed in a ``typical'' large-scale TREC evaluation (in our case, the TREC 2024 RAG Track).
To reduce the possibility of data contamination as much as possible, the test topics were publicly released only a few weeks before the evaluation deadline.
This prospective deployment provides a fair comparison between the ``standard'' (fully manual) NIST relevance assessment process and three alternative approaches that involve LLMs to varying degrees.
\citet{Soboroff:2409.15133:2024} advocated against the fully automatic approach, but entertained the human-in-the-loop processes as possible steps forward.
We let empirical analyses have the last word.

Our work, of course, does not appear in a vacuum.
The research community has known for some time now that ChatGPT (and more generally, LLMs) can be used for data labeling and other annotation tasks~\cite{Gilardi_etal_PNAS2023}, outperforming crowd-workers.
LLMs can emulate the opinions of subpopulations in answering survey questions~\cite{Chu:2303.16779:2023,KimJunsol:2305.09620:2024}, although this idea has been questioned~\cite{Dominguez-Olmedo:2306.07951:2024}. 
Another thread of work has extensively explored the use of LLMs to evaluate the output of other LLMs for a variety of tasks~\cite{Bai:2212.08073:2022,FuJinlan:2302.04166:2023,Gao:2304.02554:2023,Cohen:2305.13281:2023,ZhengLianmin:2306.05685:2023}.
This general approach is dubbed ``LLM-as-a-judge'' and represents an active area of research.
Our work builds on this vast literature, but to our knowledge, no other study like ours has ever been conducted, in terms of scale, focus on relevance assessment, and careful calibration against human-based reference baselines.

\section{Methods}

The overall structure of our study adopts the meta-evaluation methodology common in the information retrieval literature dating back several decades~\cite{Zobel98,Voorhees_SIGIR1998,Buckley00,Buckley04,Sanderson05,Harman_2011}.
Existing NIST processes for pooling and relevance assessment, refined over several decades, serve as the reference point of comparison.
System rankings generated by relevance judgments from NIST assessors are acknowledged by the community as representing the ``gold standard'' and provide a point of calibration.
This is worth emphasizing:\ unlike in prior studies dependent on crowd-sourced workers of questionable quality, our human ``baselines'' can be considered highly reliable.

We explore three different approaches using LLMs to generate relevance assessments with our \umb tool, detailed in Section~\ref{section:methods:llm-involvement}.
System rankings induced by these alternatives are compared against those induced by the NIST relevance judgments.
Correlation between these system rankings, captured in terms of Kendall's $\tau$, is taken as the measure of quality.
In other words, we ask (RQ1):\
Do relevance judgments derived from \umb lead to the same conclusions about the quality of the systems?

Of course, quality is balanced against cost.
Again, taking existing NIST processes as the reference, we ask (RQ2):\ Can we achieve the same level of quality (i.e., ability to rank systems in terms of effectiveness), but at lower costs?
With increasing LLM involvement, costs decrease---but what's the tradeoff with respect to quality?
Finally, beyond a narrow focus on just rank correlations, we ask (RQ3):\ Are there any systematic differences between human and \umb judgments?
We seek to find out the answers to all these research questions.

\subsection{TREC 2024 RAG Track}

The context for this work is the Retrieval-Augmented Generation (RAG) Track at TREC 2024.
The track was divided into three tasks:\ retrieval, augmented generation, and full RAG.
Here, we use data only from the retrieval task, which adopts a standard {\it ad hoc} setup.
Systems are presented with queries (called topics in TREC parlance), and their task is to return ranked lists of relevant passages from the MS MARCO V2.1 deduped segment collection~\cite{ragnarok}, which traces its lineage back to the original MS MARCO dataset~\cite{Bajaj:1611.09268:2018}.

\begin{figure}
\vspace{0.25cm}
\begin{itemize}
\item what is vicarious trauma and how can it be coped with? \\
\item how did the northwest coast people develop and use animal imagery in their homes? \\
\item why disability insurance is a smart investment \\
\item how bad did the vietnam war devastate the economy in 1968 \\
\item what target stors's policies for shoplifting \\
\end{itemize}
    \caption{The first five topics from the TREC 2024 RAG Track.}
    \label{fig:sample-topics}
\end{figure}

The MS MARCO V2.1 deduped segment collection~\cite{ragnarok} contains 113,520,750 text passages, derived from a deduplicated version of the MS MARCO V2 document collection~\cite{craswell2022overview}.
This underlying document collection was first cleaned by removing near-duplicates using locality sensitive hashing (LSH) with MinHash and 9-gram shingles, reducing the original document count from 11,959,635 to 10,960,555 documents. 
From this cleaned document base, passages (segments) were created using a sliding window technique with overlap---specifically, using windows of 10 sentences with a stride of 5 sentences, producing passages typically between 500--1000 characters.
To be clear, the passages represent the basic unit of retrieval, but following common IR parlance, we interchangeably refer to these as ``documents'' in a generic sense, especially when it is clear from context what we mean.

For the TREC 2024 RAG Track topics, we used a fresh scrape from Bing Search logs containing non-factoid topics warranting long-form answers that are often multifaceted and subjective~\cite{Rosset:2402.17896:2024}. 
We gathered the topics close to the submission period (July 2024) to ensure a fair evaluation when using commercial LLMs by minimizing any potential data leakage.
Here, we are most concerned about contamination with respect to relevance judgments.
Since we are retrieving from web corpora, it is highly likely that our passages have already been included in LLM pretraining data.

After some manual filtering by NIST annotators, we provided participants with a total of 301 topics.
The first five topics are shown in Figure~\ref{fig:sample-topics}.
Note that real-world information needs are often phrased awkwardly and contain misspellings; we intentionally made no attempt to fix these issues.
The end-to-end evaluation involved RAG answers to these topics, but here we only examine the retrieval stage.

Consistent with previous iterations of the TREC Deep Learning Track~\cite{craswell2022overview,Craswell_etal_TREC2023}, our evaluation used the following relevance grades, with the associated descriptions:

\begin{itemize}[leftmargin=0.5cm]
\item [0] {\bf Non-relevant}: the passage has nothing to do with the query
\item [1] {\bf Related}: the passage seems related to the query but does not contain any part of an answer to it
\item [2] {\bf Highly relevant}: the passage contains a partial or complete answer for the query, but the answer may be a bit unclear, or hidden amongst extraneous information
\item [3] {\bf Perfectly relevant}: the passage is dedicated to the query and contains a partial or complete answer to it
\end{itemize}

\noindent At a high level, the retrieval task in TREC 2024 RAG can be viewed as a natural continuation of the previous TREC Deep Learning Tracks.
Unless otherwise specified, our evaluation methodology follows what was laid out previously in those tracks.

\begin{figure}[t]
\tiny
\justifying
\begin{minted}[fontsize=\footnotesize, frame=lines, frame=single,linenos=false,breaklines,breaksymbol=,escapeinside=||,bgcolor=LightGray]{text}
Given a query and a passage, you must provide a score on an integer scale of 0 to 3 with the following meaning:
0 = represent that the passage has nothing to do with the query, 
1 = represents that the passage seems related to the query but does not contain any part of an answer to it, 
2 = represents that the passage contains a partial or complete answer for the query, but the answer may be a bit unclear, or hidden amongst extraneous information and 
3 = represents that the passage is dedicated to the query and contains a partial or complete answer to it.

Query: {query}
Passage: {passage}

Split this problem into steps:
Consider the underlying intent of the search.
Measure how well the content matches a likely intent of the query (M).
Measure how trustworthy the passage is (T).
Consider the aspects above and the relative importance of each, and decide on a final score (O). Final score must be an integer value only.
Do not provide any code in result. Provide each score in the format of: ##final score: score without providing any reasoning.
\end{minted}
\caption{The prompt utilized with \umb for relevance assessment.}
\label{fig:prompt}
\end{figure}

\begin{table*}[t]
  \begin{tabular}{ll | c  c | r r r r | c}
    \toprule
    & \multirow{2}{*}{\textbf{Condition}} & \multirow{2}{*}{\textbf{\# topics}} & \textbf{avg pool} & \multicolumn{4}{c}{\textbf{avg \# docs assessed}} & \textbf{avg pool size} \\
    &&& \textbf{size} & \textbf{~ ~ 0} & \textbf{~ ~ 1} & \textbf{~ ~ 2} & \textbf{~ ~ 3} & \textbf{decrease} \\
    \midrule
    \midrule
    (1a) & fully manual & 27 & 332 & 172 & 73 & 57 & 27 & \\
    (1b) & manual w/ filtering & 29 & 188 & 52 & 67 & 54 & 15 & \\
    (1c) & manual w/ post-editing & 30 & 195 & 46 & 79 & 50 & 19 & \\
    \midrule
    (2a) & \umb (for fully manual) & 27 & 332 & 108 & 115 & 77 & 33 & \\
    (2b) & \umb (for manual w/ filtering) & 29 & 305 & 117 & 121 & 51 & 17 & $-$38\% \\
    (2c) & \umb (for manual w/ post-editing) & 30 & 332 & 137 & 124 & 51 & 20 & $-$41\% \\
    \midrule
    (3) & \umb (overall) & 301 & 360 & 132 & 143 & 65 & 20 & \\
  \bottomrule
  \end{tabular}
  \vspace{0.25cm}
  \caption{Descriptive statistics for relevance judgments under different conditions. Note that (a), (b), and (c) are disjoint.}
  \label{tab:qrel_stats}
\end{table*}

\subsection{UMBRELA}

We used \umb~\cite{Upadhyay_etal_arXiv2024} to generate automatic relevance labels in this work.
As discussed in Section~\ref{section:background}, the tool is an open-source reproduction of the work described in~\citet{ThomasPaul_etal_SIGIR2024}, utilizing the DNA prompting technique described by the authors.

Figure~\ref{fig:prompt} shows the exact prompt used in \umb for relevance assessment.
Similar to the details provided in \citet{Upadhyay_etal_arXiv2024}, we employed zero-shot prompting to perform relevance assessment using OpenAI's  GPT-4o model.
Note that we used the publicly available version, as opposed to a private variant used in \citet{ThomasPaul_etal_SIGIR2024}.
\umb leverages precise instructions to assist the LLM in understanding the semantic relation between the query and the passage and thus guides the LLM in assigning (hopefully) accurate labels.

\subsection{Level of LLM Involvement}
\label{section:methods:llm-involvement}

We explored four different approaches to relevance assessments, detailed below.
Common to all these approaches is the construction of the pools to depth 20 from the run submissions to the TREC 2024 RAG Track.

\paragraph{Fully Manual Assessment}
This condition represents the existing NIST methodology for relevance assessment.
Operationally, the deployed processes were taken from the TREC 2023 Deep Learning Track~\cite{Craswell_etal_TREC2023}, although the overall workflow dates back to the {\it ad hoc} document retrieval tasks from the 1990s (with various refinements over the years).


\paragraph{Manual Assessment with Filtering}
In this condition, \umb was used to filter the pools, discarding passages labelled as non-relevant by the model.
That is, such passages were not shown to the NIST assessor.
Importantly, however, the assessors did {\it not} have access to the \umb judgments.
Operationally, this condition is indistinguishable from fully manual assessment; the assessors were just presented with different (smaller) pools.


\paragraph{Manual Post-Editing of Automatic Assessment}
In this condition, we used \umb to assess the pools, and after discarding passages judged to be non-relevant, we presented the remaining to the assessors.
Here, the assessors were made aware of the \umb labels.
In other words, the NIST assessors were post-editing LLM judgments.
This was operationally performed by grouping all passages of the same relevance grade together and explicitly informing the assessor.
That is, the assessor was told (for example), ``Here are all the passages that the LLM found to be highly relevant''.


\paragraph{Fully Automatic Assessment}
In this condition, we used relevance labels from \umb, unmodified, to evaluate runs directly.
This required no manual involvement and thus we were able to assess all evaluation topics.


\section{Results}

\subsection{Descriptive Statistics}
\label{section:results:stats}

For the TREC 2024 RAG Track, we received a total of 77 retrieval-only runs from 19 teams.
Of these, 5 runs were from the organizers, who contributed a range of baselines; these are included in the statistics provided above.
All participants were given the set of 301 topics and were requested to return ranked lists for all of them.

Descriptive statistics for the relevance assessments are shown in Table~\ref{tab:qrel_stats}, where we break down statistics into the various conditions described in Section~\ref{section:methods:llm-involvement}.
NIST assessors were given the option to judge whatever topics interested them, selecting from the set of 301 topics.
In total, they were able to assess 27 topics using the fully manual process, 31 topics for manual with filtering, and 31 topics for manual with post-editing.
However, for 3 topics, NIST assessors found no passages with at least a relevance grade of 1 (i.e., ``related''):\ two of these occurred in the manual with filtering condition and one in the manual with post-editing condition.
Thus, these topics were removed from consideration, leading to the final figures presented in Table~\ref{tab:qrel_stats}, shown in rows (1a), (1b), and (1c).
To be clear, the topics for each of these conditions are disjoint.
Since \umb is fully automatic, we were able to apply it to all 301 topics in the test set, whose statistics are shown in row (3).

In Table~\ref{tab:qrel_stats}, the row (2) entries provide descriptive statistics for the common (i.e., overlapping) topics between the \umb condition and each of the other conditions.
That is, (2a) provides statistics from the \umb assessments that correspond to the 27 topics that received fully manual assessments, i.e., row (1a).
The same correspondence holds between (2b) and (1b), and between (2c) and (1c).
By comparing row (2b) vs.\ row (1b) and row (2c) vs.\ row (1c), we can quantify the amount of effort saved from using an LLM, since in both cases, passages assessed as non-relevant by \umb are not shown to the assessors.
This is presented in the last column, where \umb reduces the assessment pool by roughly 40\%.

From these results, we see differences in distributions across the relevance grades.
Focusing on the LLM-assisted approaches, row (2b) vs.\ (1b) and row (2c) vs.\ (1c), we see more passages assessed as ``related''.
However, there does not appear to be obvious differences in the higher grades, ``highly relevant'' and ``perfectly relevant''.
The differences between row (2a) vs.\ (1a) suggest that \umb adopts more lenient relevance criteria, leading to more passages at each of the 1, 2, and 3 grades, compared to fully manual assessments.

 \begin{figure*}[bth]
        \centering
        \includegraphics[width=0.32\textwidth]{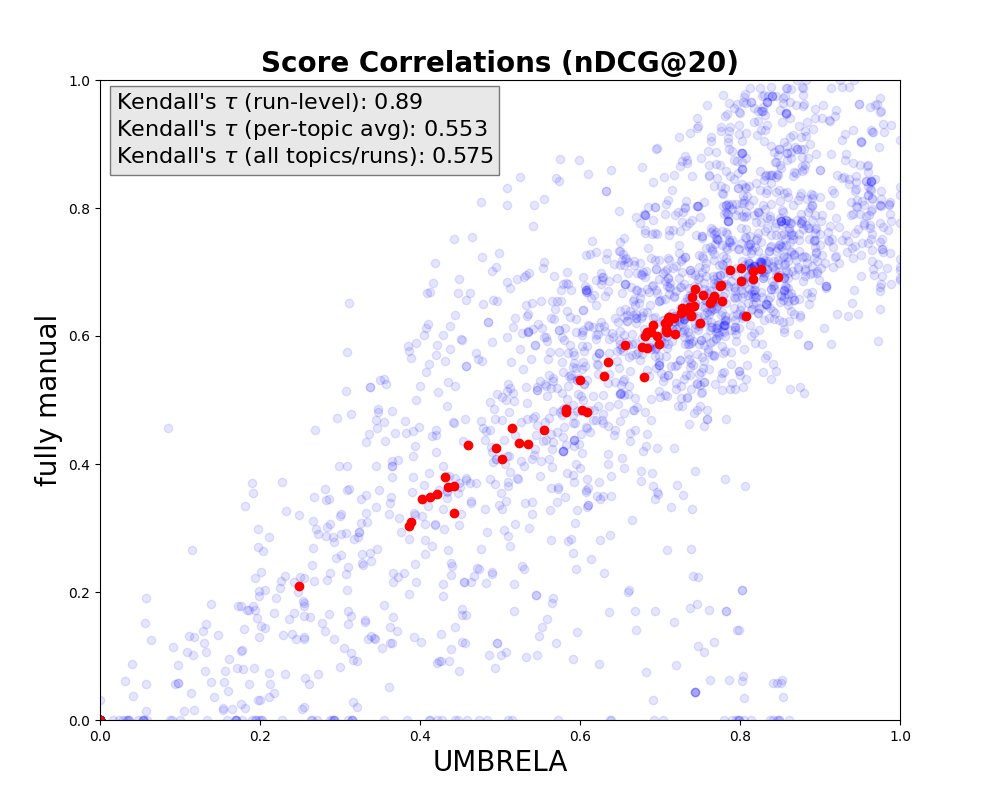}
        \includegraphics[width=0.32\textwidth]{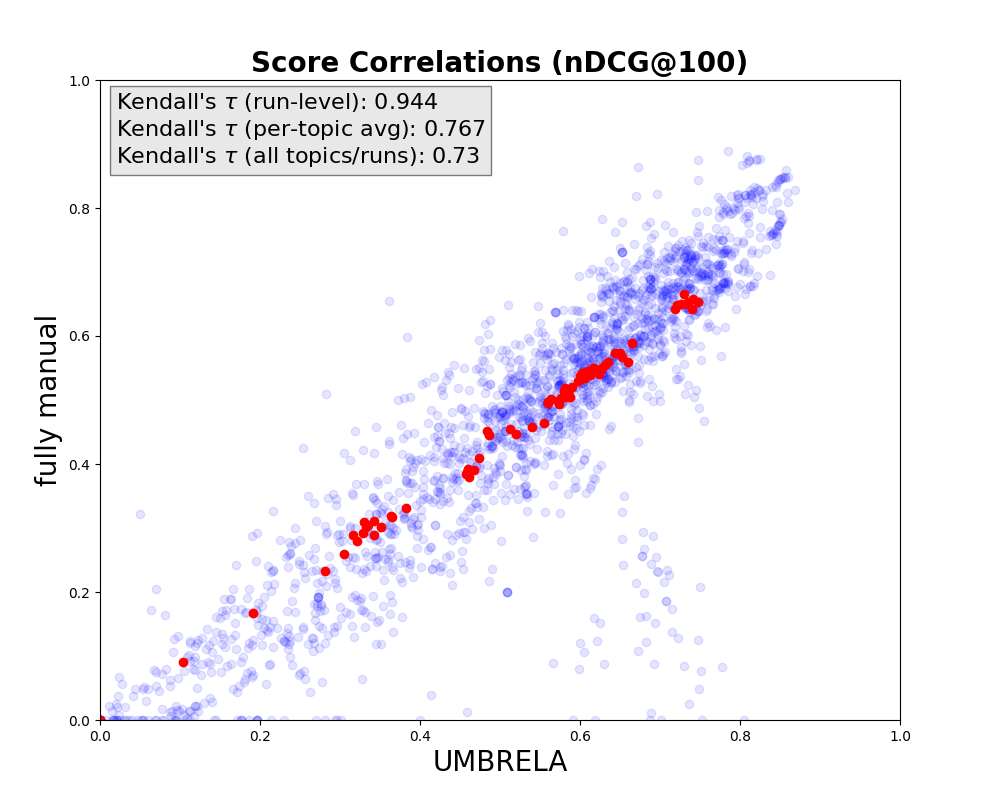}
        \includegraphics[width=0.32\textwidth]{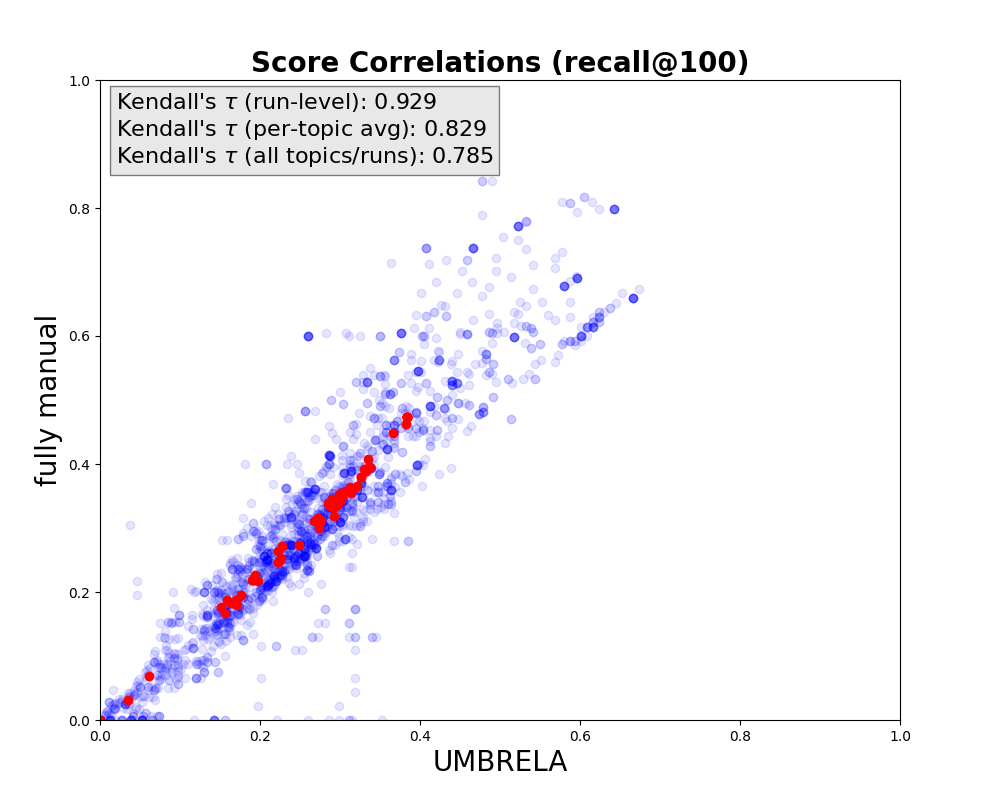} \\
        \includegraphics[width=0.32\textwidth]{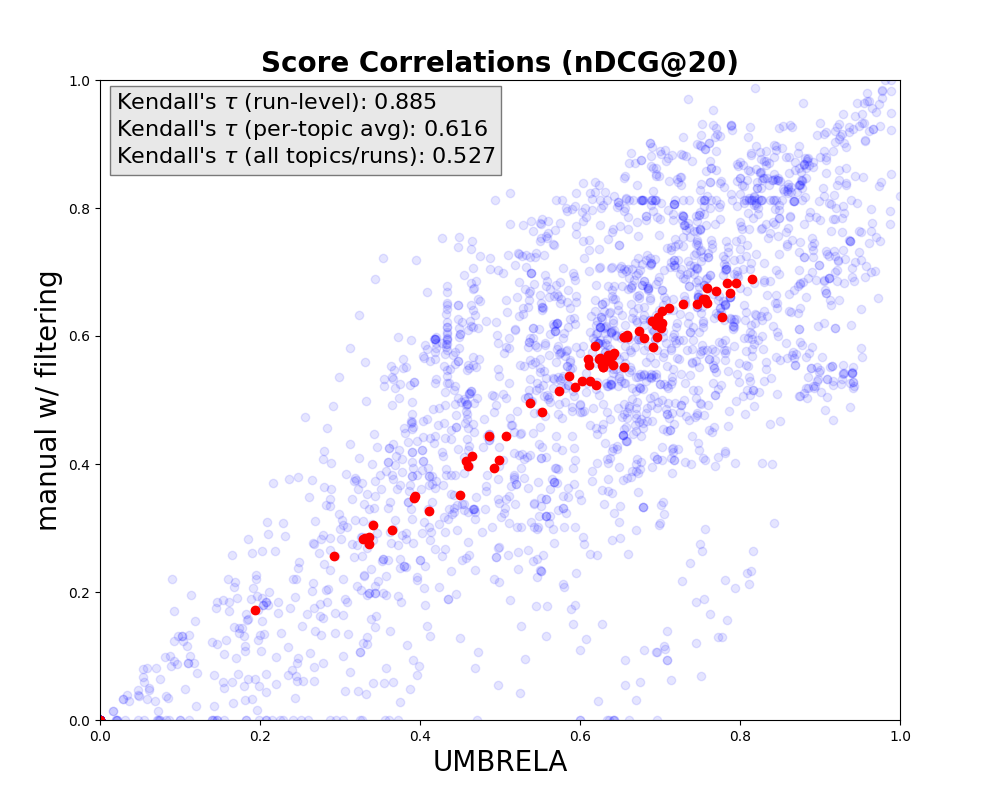}
        \includegraphics[width=0.32\textwidth]{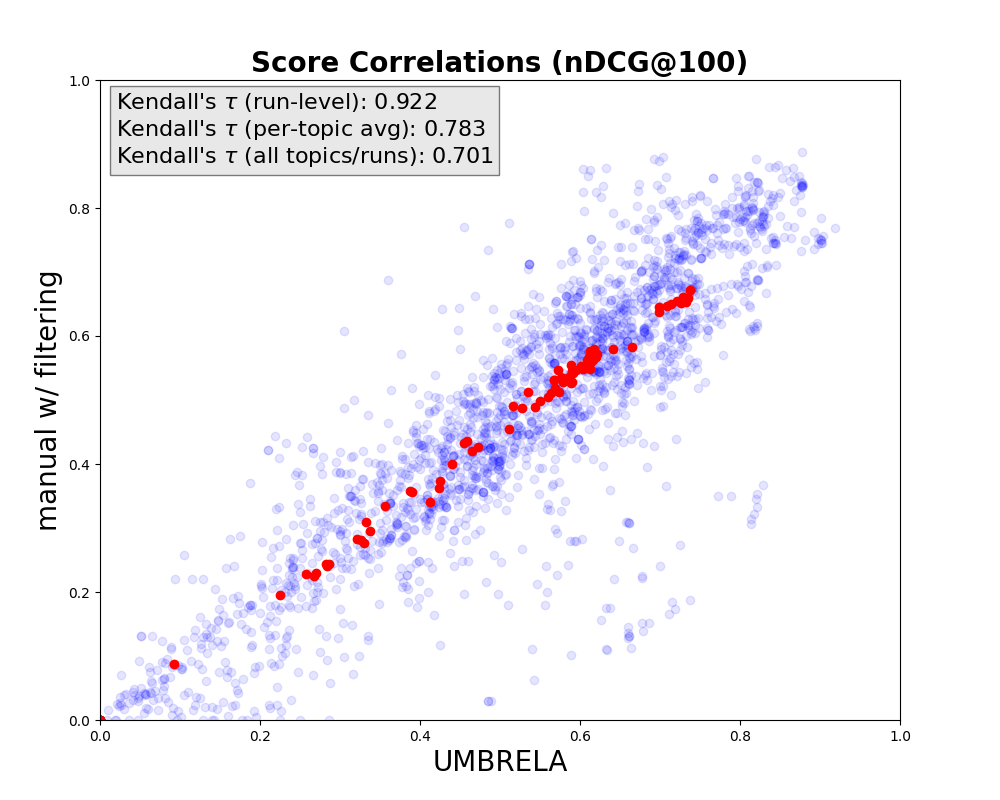}
        \includegraphics[width=0.32\textwidth]{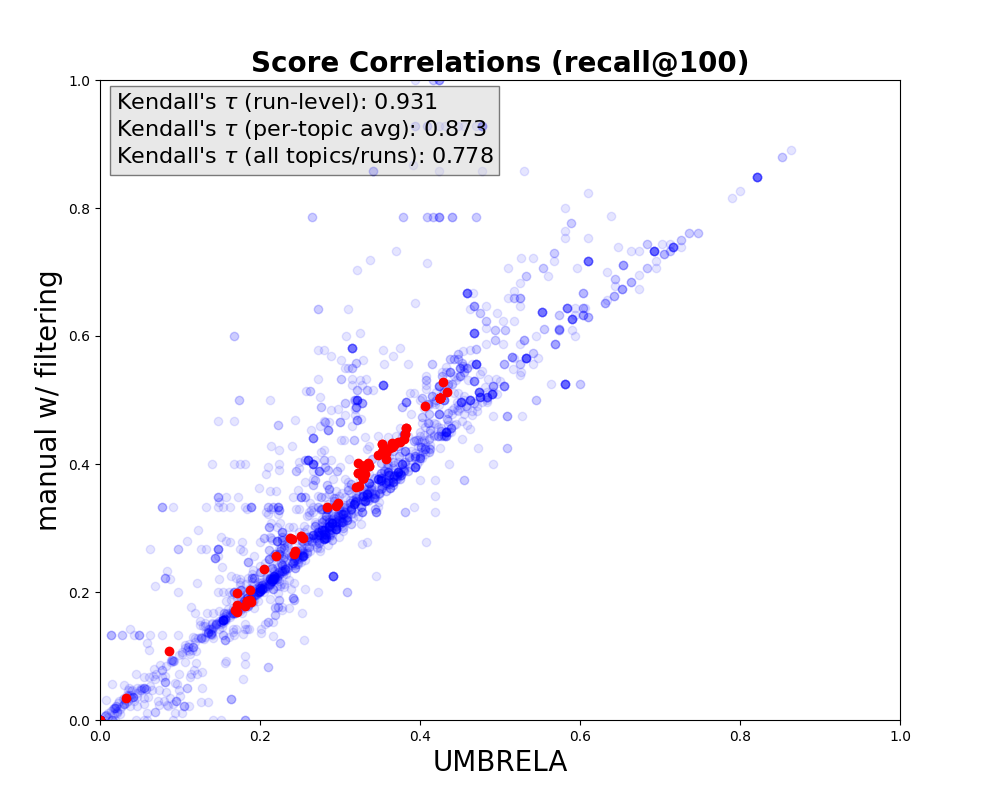} \\
        \includegraphics[width=0.32\textwidth]{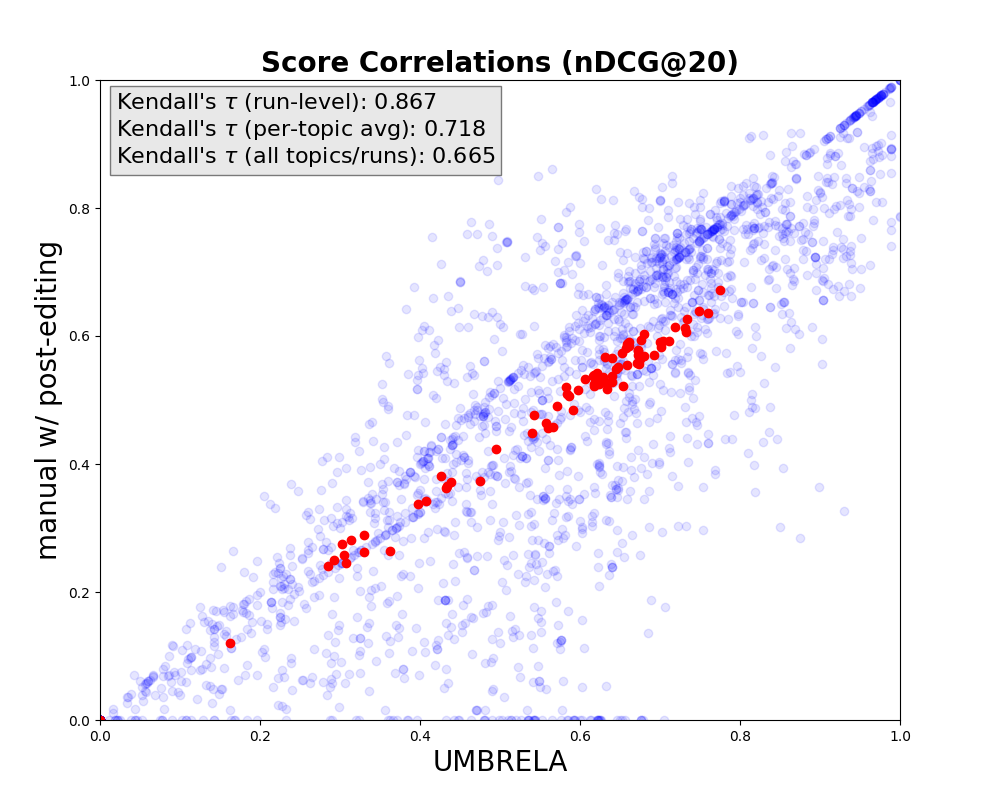}
        \includegraphics[width=0.32\textwidth]{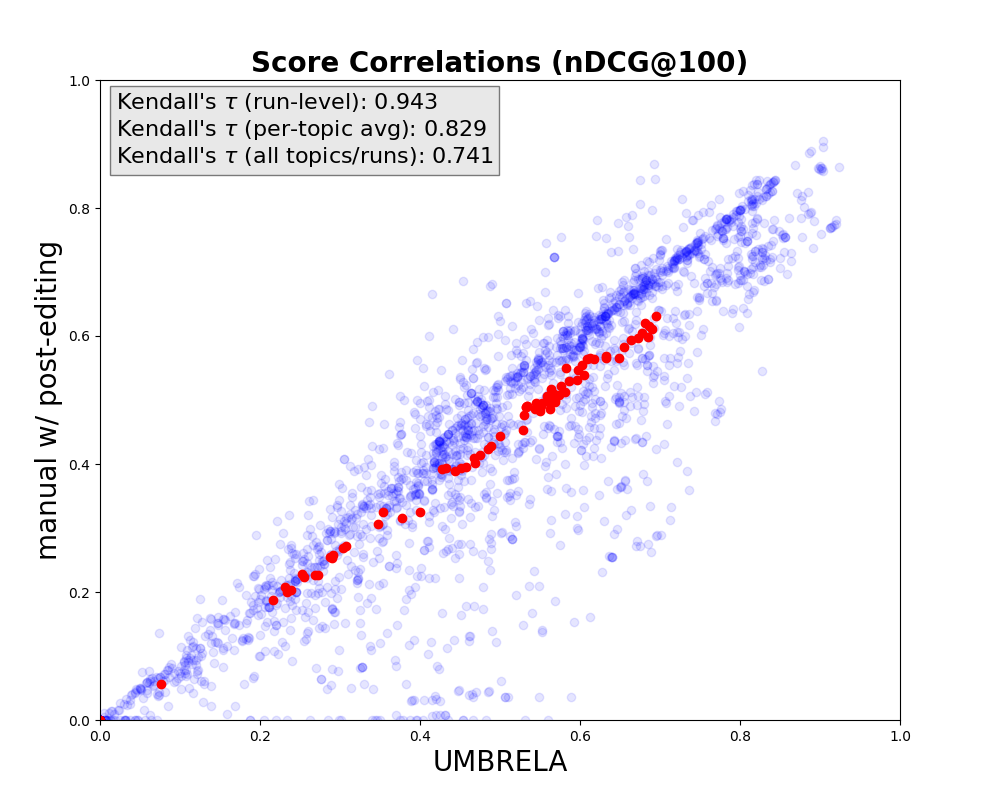}
        \includegraphics[width=0.32\textwidth]{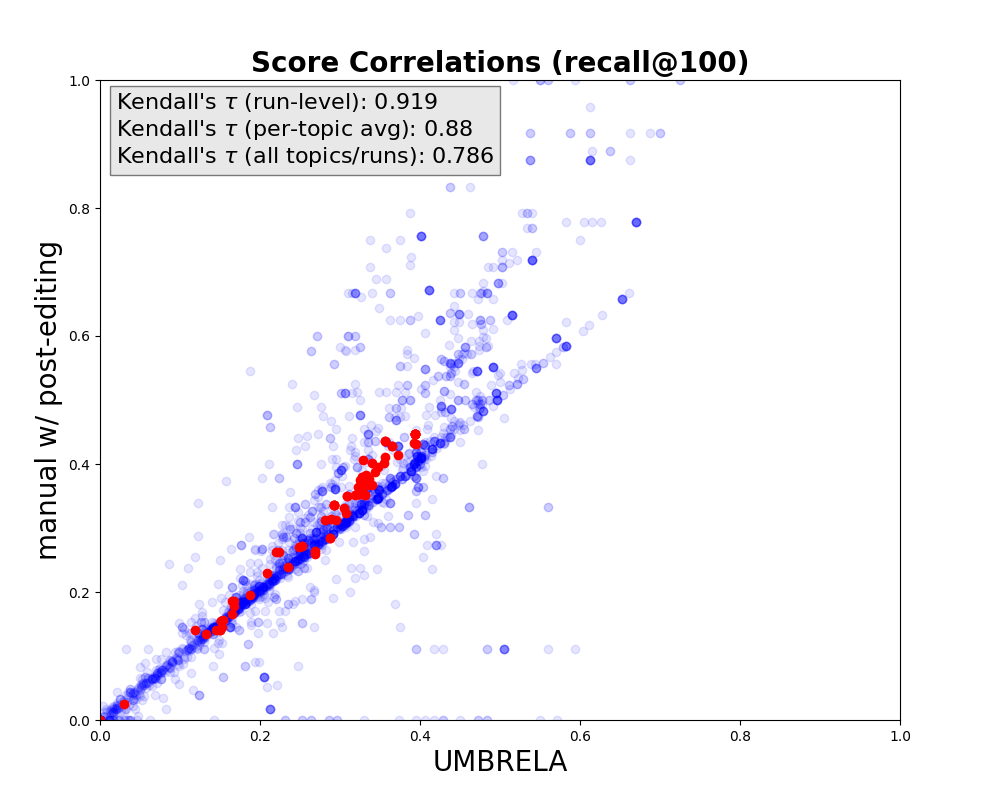}
    \caption{Comparisons between \umb scores and scores from fully manual (top row), manual with filtering (middle row), and manual with post-editing (bottom row). Columns show different metrics:\ nDCG@20, nDCG@100, and Recall@100. In each scatter plot, red dots show run-level scores and the blue dots show all topic/run combinations. Each scatter plot is annotated with rank correlations in terms of Kendall's $\tau$. This analysis is performed on common (i.e., overlapping) topics.}
    \label{fig:scatter_plots_1}
\end{figure*}

\subsection{Rank Correlations}

\begin{table*}[t]
  \begin{tabular}{ll|ll|ll|ll}
    \toprule
    & \textbf{Condition} & \multicolumn{2}{c}{\textbf{nDCG@20}} & \multicolumn{2}{c}{\textbf{nDCG@100}} & \multicolumn{2}{c}{\textbf{Recall@100}} \\
    & \umb vs. & Run-Level & Per-Topic & Run-Level & Per-Topic & Run-Level & Per-Topic \\
    \midrule
    \midrule
    (1a) & fully manual & 0.890 & 0.553 & 0.944 & 0.767 & 0.929 & 0.829 \\
    (1b) & fully manual (2/3) & 0.811 $\pm$ 0.06 & 0.544 $\pm$ 0.09 & 0.904 $\pm$ 0.04 & 0.762 $\pm$ 0.09 & 0.903 $\pm$ 0.05 & 0.826 $\pm$ 0.07 \\
    (1c) & fully manual (1/3) & 0.768 $\pm$ 0.08 & 0.566 $\pm$ 0.12 & 0.875 $\pm$ 0.08 & 0.777 $\pm$ 0.10 & 0.878 $\pm$ 0.07 & 0.841 $\pm$ 0.08 \\
    \hline
    (2a) & manual w/ filtering & 0.885 & 0.616 & 0.922 & 0.783 & 0.931 & 0.873 \\
    (2b) & manual w/ filtering (2/3) & 0.861 $\pm$ 0.05 & 0.615 $\pm$ 0.08 & 0.905 $\pm$ 0.04 & 0.785 $\pm$ 0.04 & 0.912 $\pm$ 0.04 & 0.874 $\pm$ 0.06 \\
    (2c) & manual w/ filtering (1/3) & 0.814 $\pm$ 0.10 & 0.611 $\pm$ 0.13 & 0.878 $\pm$ 0.07 & 0.780 $\pm$ 0.07 & 0.887 $\pm$ 0.07 & 0.873 $\pm$ 0.09 \\
    \hline
    (3a) & manual w/ post-editing & 0.867 & 0.718 & 0.943 & 0.829 & 0.919 & 0.880 \\
    (3b) & manual w/ post-editing (2/3) & 0.844 $\pm$ 0.07 & 0.719 $\pm$ 0.09 & 0.917 $\pm$ 0.04 & 0.824 $\pm$ 0.06 & 0.872 $\pm$ 0.07 & 0.876 $\pm$ 0.06 \\
    (3c) & manual w/ post-editing (1/3) & 0.814 $\pm$ 0.10 & 0.726 $\pm$ 0.14 & 0.897 $\pm$ 0.06 & 0.834 $\pm$ 0.10 & 0.862 $\pm$ 0.11 & 0.887 $\pm$ 0.10 \\
  \bottomrule

  \end{tabular}
  \vspace{0.25cm}
  \caption{Rank correlations (Kendall's $\tau$) at the run level and averaged on a per-topic basis, comparing \umb vs.\ fully manual in row (1), vs.\ manual with filtering in row (2), vs.\ manual with post-editing in row (3). Values in rows (*a) are exactly those shown in Figure~\ref{fig:scatter_plots_1}. Rows (*bc) report correlations based on sampling $2/3$ and $1/3$ of topics, respectively, using a Monte Carlo simulation, with 95\% confidence intervals reported.}
  \label{tab:kendall_tau}
\end{table*}

At a high-level, our analyses compute rank correlations between evaluation scores induced by \umb judgments (qrels) vs.\ judgments (qrels) derived by the other assessment processes.
Following common practice in IR meta-evaluations, rank correlation is captured using Kendall's $\tau$.
However, there are multiple ways to design a rank correlation analysis.
In the following, we explain the options in terms of \umb vs.\ fully manual assessments, but we apply the same methodology to the other conditions:

\begin{itemize}[leftmargin=*]

\item Run-level correlations on common topics (run-level).
For each run, we evaluate with qrels from the 27 topics with fully manual judgments and \umb qrels for the same 27 topics.
These correspond to row (1a) and row (2a) in Table~\ref{tab:qrel_stats}, respectively.
We computed Kendall's $\tau$ based on these two scores.

\item Average of per-topic correlations (per-topic avg).
Same as above:\ For each run, we evaluate with qrels from the 27 topics with fully manual judgments and \umb qrels for the same 27 topics.
These correspond to row (1a) and row (2a) in Table~\ref{tab:qrel_stats}, respectively.
However, unlike above, we compute Kendall's $\tau$ {\it for each topic}, and then average across the per-topic correlations.

\item Correlation across all topic/run combinations (all topics/runs).
We evaluate with the same qrels as above, but each topic/run combination is considered an independent observation, and we computed Kendall's $\tau$ across all these observations.

\end{itemize}

\noindent Another important aspect of experimental design is the effectiveness metric.
We focus on three:\ nDCG@20, nDCG@100, and Recall@100.
The first is perhaps the most common as it evaluates early precision, representing a common setting in RAG applications; i.e., top-20 results are fed into the LLM prompt for answer generation.
However, the downside of nDCG@20 is that most topics have more than 20 relevant passages (see Table~\ref{tab:qrel_stats}), in which case nDCG@100 better captures effectiveness.
Recall is typically used to quantify the upper bound of effectiveness in a multi-stage pipeline, and a cutoff at 100 passages represents a typical setting.

Results of our analyses are shown in Figure~\ref{fig:scatter_plots_1} as scatter plots:\ the left column shows nDCG@20 scores, the middle column shows nDCG@100 scores, and the right column shows Recall@100 scores.
The top row compares \umb with fully manual judgments; the middle row, manual with filtering, and the bottom row, manual with post-editing.
In each plot, the red solid dots represent run-level scores (simple average across all topics, per standard IR practice) and the blue dots show all topic/run combinations.
In other words, the correlation between the red dots captures run-level correlations on common topics and the correlation between the blue dots capture correlation across all topic/run combinations.
The average of per-topic correlations is not readily visualized on these plots.

In the discussion below, we introduce the notion of ``scatter'' to describe score disagreements across runs and topics.
Perfect agreement (i.e., no scatter) would have all run/topic combinations (i.e., the blue dots in the plots) lie on a straight line.
We can quantify scatter in terms of Kendall's $\tau$ on all topic/run observations (i.e., the blue dots):\ the lower the scatter, the greater the correlation.

In terms of nDCG@20 (left column in Figure~\ref{fig:scatter_plots_1}), we see a high degree of correlation between scores induced by \umb and the other conditions, i.e., the red dots.
We do see a fair amount of scatter for the blue dots, indicating disagreement across individual run/topic observations.
In the manual with post-editing case (bottom left), there appears to be less scatter, as the assessors are provided an anchor in terms of the \umb judgments (but are free to change them).
Interestingly, there does not appear to be a noticeable difference between the fully manual assessments (top) and the other two LLM-assisted alternatives (middle and bottom) in terms of run-level rank correlations.

The conclusions are similar for nDCG@100 (middle column) and Recall@100 (right column), with slightly higher rank correlations than their nDCG@20 counterparts.
We observe less scatter for nDCG@100 and Recall@100 across all topic/run observations (i.e., the blue dots) compared to nDCG@20, which makes sense since evaluating deeper ranked lists should improve metric stability.
As with nDCG@20, the two LLM-assisted processes do not exhibit noticeably higher rank correlations with \umb.

\smallskip \noindent So, to provide an answer to RQ1:\

\begin{itemize}

\item[{\bf RQ1}] Automatically generated \umb judgments can replace fully manual assessments for common effectiveness metrics at the run level (in the context of the TREC-style evaluations).

\end{itemize}

\noindent Overall, this finding is consistent with results from the Workshop on Large Language Models (LLMs) for Evaluation in Information Retrieval at SIGIR 2024~\cite{Rahmani:2408.08896:2024}.

The plots in Figure~\ref{fig:scatter_plots_1} show that despite variance across topic/run combinations (i.e., the scatter of the blue dots), we observe good rank correlations after aggregating scores across topics ($\sim$30 in our case).
A natural question is:\ How many topics do we need?

\begin{figure*}[bth]
        \centering
        \includegraphics[width=0.32\textwidth]{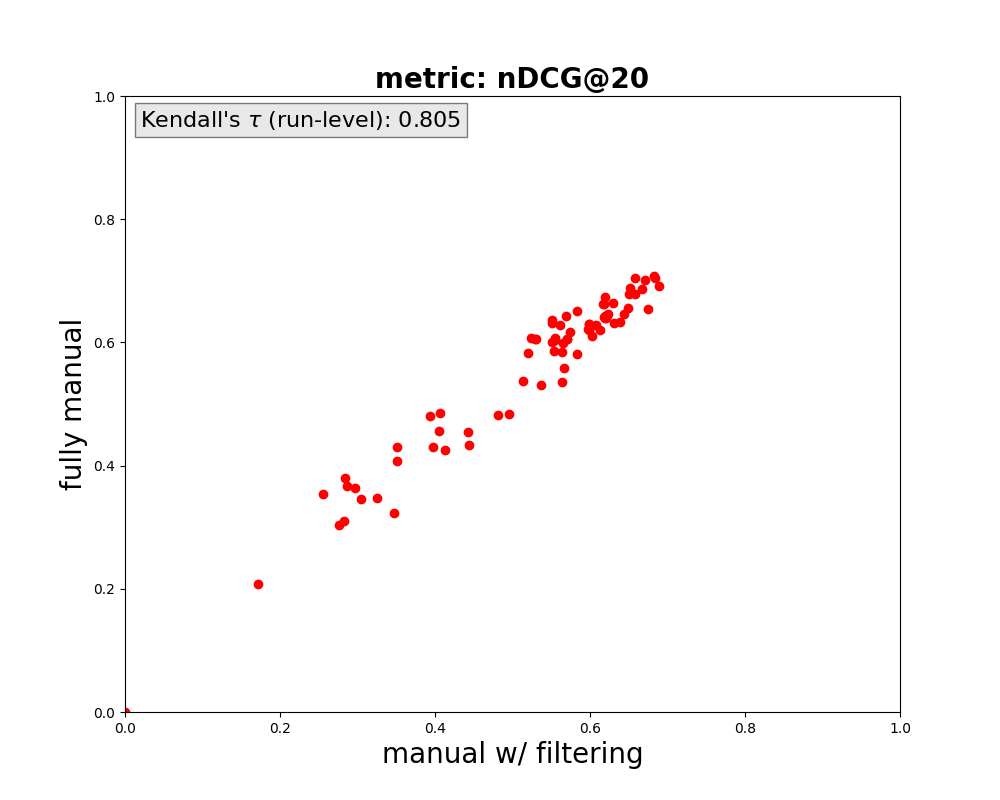}
        \includegraphics[width=0.32\textwidth]{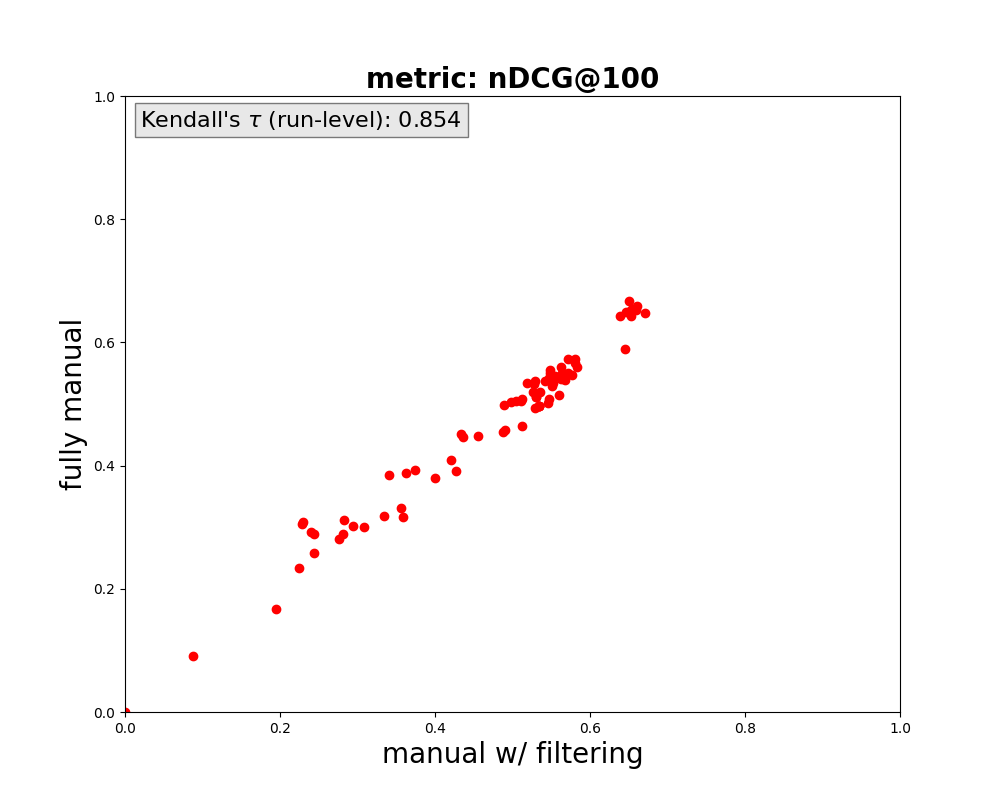}
        \includegraphics[width=0.32\textwidth]{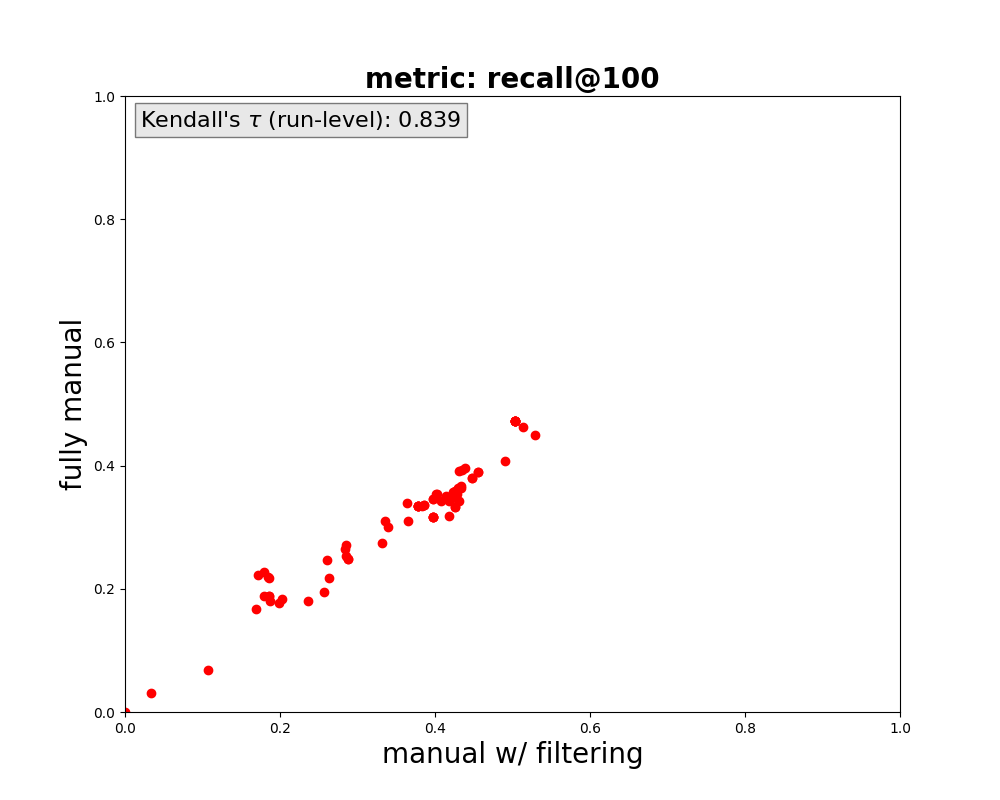} \\
        \includegraphics[width=0.32\textwidth]{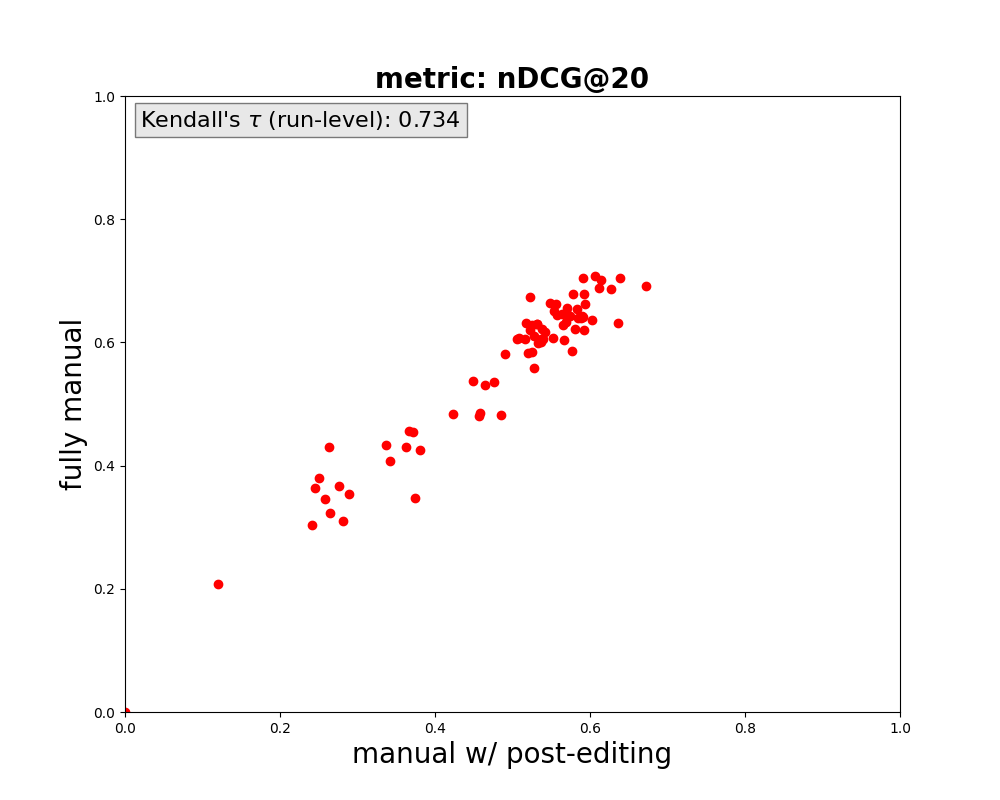}
        \includegraphics[width=0.32\textwidth]{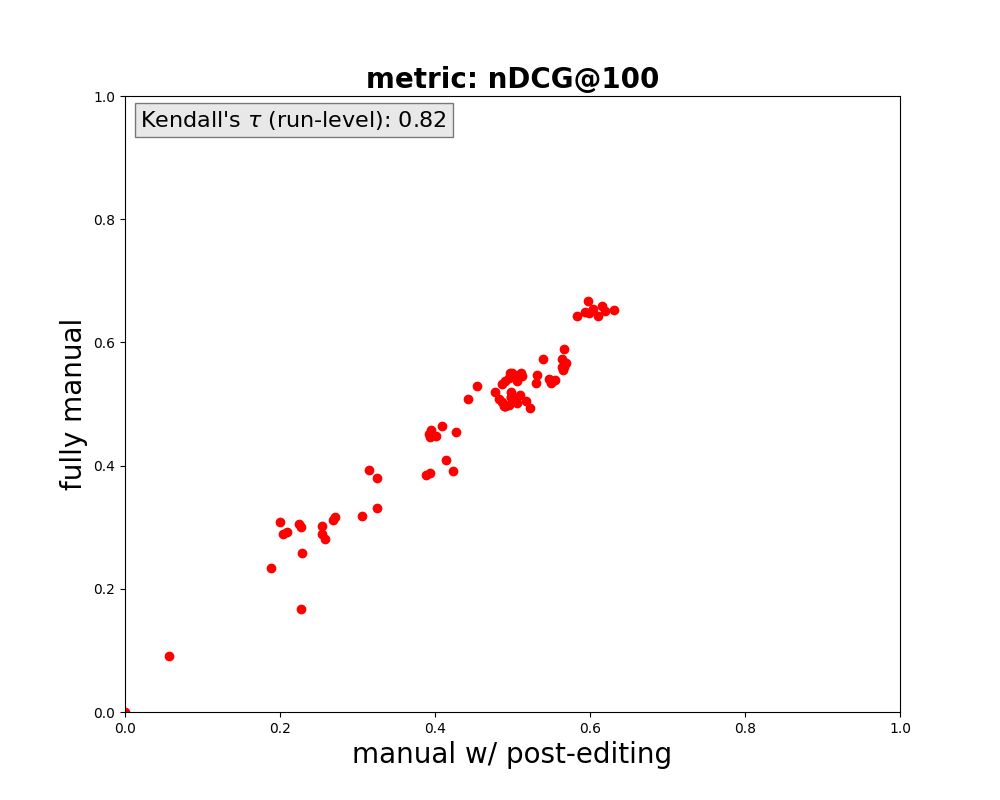}
        \includegraphics[width=0.32\textwidth]{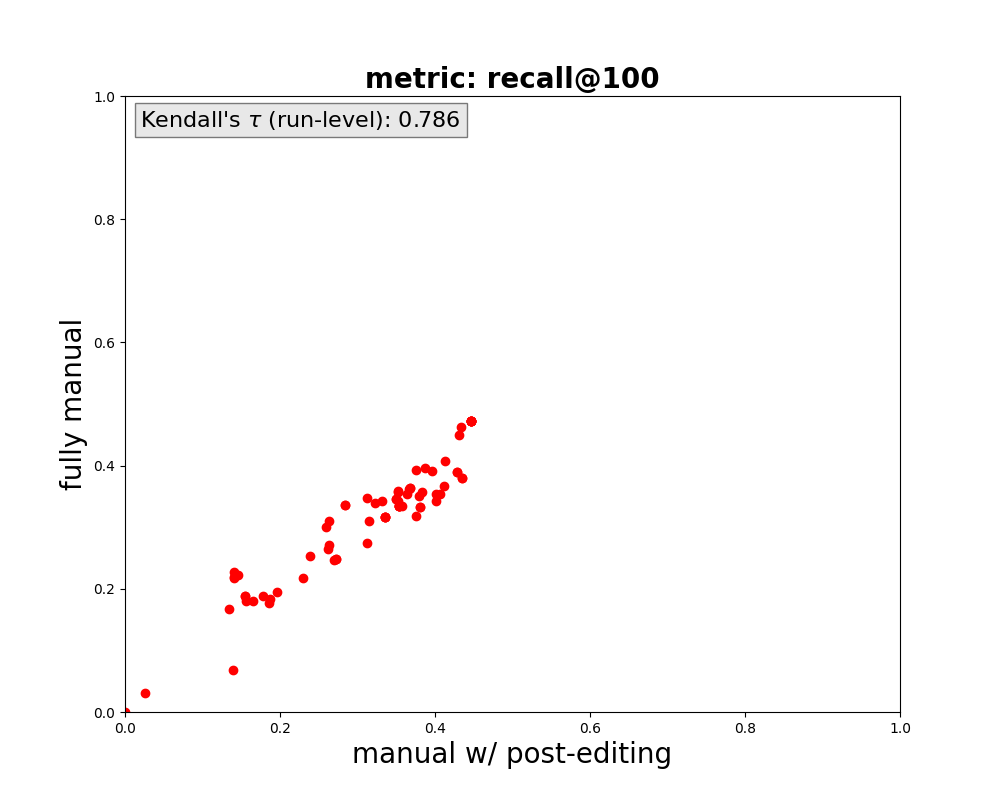}
    \caption{Run-level rank correlations (Kendall's $\tau$) comparing manual with filtering vs.\ fully manual (top row) and manual with post-editing vs.\ fully manual (bottom row). Columns show different metrics:\ nDCG@20, nDCG@100, and Recall@100. Note that topics are disjoint in this analysis.}
    \label{fig:scatter_plots_2}
\end{figure*}

The answer can be found in Table~\ref{tab:kendall_tau}, where row (1a), (2a), and (3a) simply repeat the Kendall's $\tau$ correlation figures from Figure~\ref{fig:scatter_plots_1}.
The other rows repeat exactly the same procedure for computing rank correlation, but over (roughly) thirds of the topics using a Monte Carlo simulation.
For example, in row (2b), we compute Kendall's $\tau$ over $2/3$ of the topics (random sampling without replacement) in the fully manual condition (i.e., 27 choose 18).
We repeat the process 100 times, from which we can compute the mean and confidence intervals across all the trials.
Here, the 95\% confidence interval is determined by taking the 2.5\% and 97.5\% percentiles of the means computed over the trials.
In row (1c), we repeat, but with only $1/3$ of the topics; the (b) and (c) rows of (2) and (3) represent the corresponding conditions with the LLM-assisted processes.
As expected, run-level rank correlations decrease with fewer topics, but these results suggest that we can achieve reasonable run-level rank correlations with a few as a $\sim$10 topics.

In Figure~\ref{fig:scatter_plots_2}, we set aside \umb assessments and compare scores induced by the fully manual process ($y$-axis) with manual with filtering (top row) and manual with post-editing (bottom row).
The columns show different metrics:\ nDCG@20, nDCG@100, and Recall@100, as in Figure~\ref{fig:scatter_plots_1}.
Note that here we are evaluating scores from {\it disjoint} topics that slightly differ in quantity (e.g., 27 topics for fully manual, but 30 topics for manual with post-editing).
Because of this setup, we can only plot run-level correlations (i.e., red dots); there is no equivalent notion of each topic/run combination as independent observations (corresponding to the blue dots).

Comparing Figure~\ref{fig:scatter_plots_2} to Figure~\ref{fig:scatter_plots_1}, we do not see any obvious differences in rank correlations.
That is, if we take the fully manual assessment process as the ``gold standard'', rankings induced by \umb do not seem any worse than the rankings induced by the other LLM-assisted processes.
Phrased another way, human involvement does {\it not} appear to increase correlation with the gold standard (fully manual).
this suggests that expending human effort does not seem to yield a payoff in terms of more accurate system rankings.
Thus, to directly answer RQ2:\ 

\begin{itemize}

\item[{\bf RQ2}] The additional costs associated with hybrid human--LLM
solutions (i.e., human-in-the-loop efforts) do not appear to have obvious tangible benefits over fully automatic processes.

\end{itemize}

\noindent Phrased differently, we do not notice any obvious payoffs that justify the higher costs associated with human--LLM hybrid strategies, such as those advocated in previous work~\cite{Faggioli_etal_ICTIR2023,Soboroff:2409.15133:2024}.

\begin{figure*}[bth]
        \centering
        \includegraphics[width=0.32\textwidth]{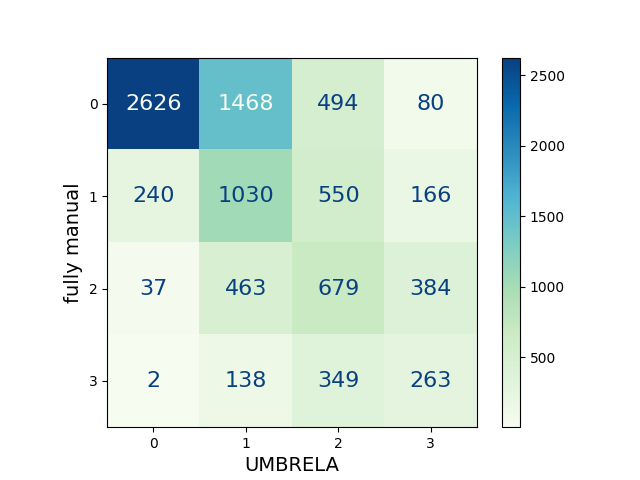}
        \includegraphics[width=0.32\textwidth]{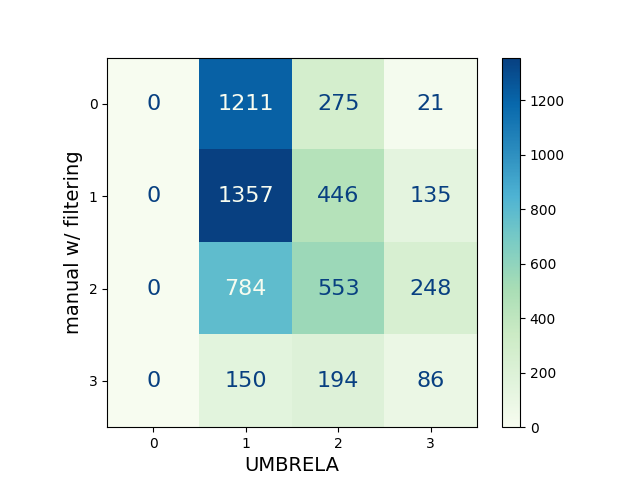} 
        \includegraphics[width=0.32\textwidth]{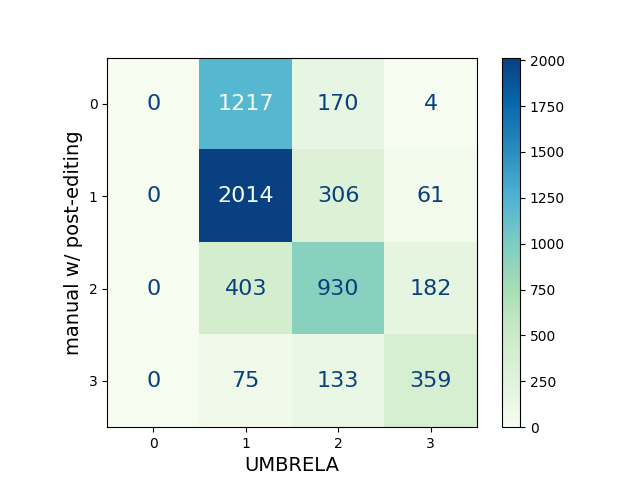}
    \caption{Confusion matrices comparing \umb with the fully manual process (left), manual with filtering (middle), and manual with post-editing (right).}
    \label{fig:conf_matrix}
\end{figure*}

\subsection{Confusion Matrices}

We next attempt to characterize differences between \umb judgments and those arising from the fully manual and LLM-assisted processes.
Confusion matrices are shown in Figure~\ref{fig:conf_matrix}:\ \umb is compared to fully manual on the left, to manual with filtering in the middle, and to manual with post-editing on the right.
While \citet{Arabzadeh:2404.04044:2024} conducted experiments that are similar in spirit to these, validating the alignment of human and LLM assessments using TREC DL 2019--20 data, their results are not directly comparable to our confusion matrices.

Consistent with the descriptive statistics in Table~\ref{tab:qrel_stats}, these results suggest that NIST assessors tend to adopt stricter relevance criteria.
For example, comparing \umb with fully manual judgments, the left panel shows that for 263 cases, both \umb and NIST assessors agreed that a document is ``perfectly relevant'' to a query.
However, there is an even greater number (384 passages) that \umb assessed as ``perfectly relevant'', but the human assessor thought was only ``highly relevant''.
We can draw similar conclusions by examining the numbers above the diagonals:\
human assessors annotate many passages as being less relevant than the labels provided by \umb.
In contrast, there are relatively fewer passages that the human assessors considered more relevant than \umb (i.e., numbers below the diagonal).
Note that in this condition, the fully manual assessment process is completely independent from \umb.

However, based on our own manual examination of the evaluation data, we can identify cases where our judgments align closer to \umb than the NIST assessors.
One example is for the query ``why are cancer rates higher on the east coast''; our interpretation is that ``east coast'' refers to the east coast of the United States.
There is a document in the pool about cancer rates in populations with ``Ashkenazi Jewish (Eastern European) heritage'', which the NIST assessor  labeled as perfectly relevant, but \umb labeled as not relevant.
Perhaps this divergence stems from a different interpretation of the topic (or perhaps this is simply an ``erroneous'' judgment), but here we agree with \umb.

Next, consider the LLM-assisted conditions.
In the middle panel and right panel in Figure~\ref{fig:conf_matrix}, the leftmost columns are empty because the NIST assessors are never shown passages judged non-relevant by \umb.
Recall that the only difference between the two conditions is that with post-editing, the NIST assessor is shown the \umb relevance grade to provide an anchor point of reference.
Thus, in the rightmost confusion matrix, where we can interpret each column as the NIST assessor either ``downgrading'' or ``upgrading'' \umb assessments (since the human assessor is provided the \umb label).
For example, in the final (``3'') column, NIST assessors ``downgraded'' 182 passages from ``perfectly relevant'' to ``highly relevant'', and 4 passages all the way down to not relevant.

Looking at the LLM-assisted conditions, we provide a few interesting case studies:
For the topic ``what change did marius make in the roman empire'', we encounter a document written as a series of ``clues'', e.g., ``He was a member of the lower-class of Romans who rose to become commander of the army.''
Although this document does {\it not} explicitly reference Marius, the clues describe him accurately.
\umb marked this document as ``perfectly relevant'', having made the correct inference, but the NIST assessor marked the document as non-relevant.

There are also cases where differences can be attributed to topic interpretation.
Consider the topic ``what steps should individuals take to dispose of their electronic devices?''
A document that discusses recycling batteries and light bulbs was marked perfectly relevant by \umb but was considered not relevant by the NIST assessor.
We suspect that the difference arises from whether batteries and light bulbs should be considered ``electronic devices''; here, \umb takes a more expansive view.
Again, we might consider this as another example of \umb making inferences that humans would find less warranted. 


Interestingly, from the rightmost confusion matrix in Figure~\ref{fig:conf_matrix}, for 75 passages that \umb labeled as only ``related'', the human assessor ``upgraded'' the document to be ``perfectly relevant''.
In at least some of these cases, it appears that the NIST assessor made an inference that \umb either didn't make or missed.
For the query ``what society issues did goth rock address'', we found a document that mentions goth in passing, but the connection was not picked up by \umb.
Overall, the cases where NIST assessors found a document to be more relevant than \umb are relatively rare, compared to the other way around.


Our analysis is missing a comparison between two {\it human} assessors, since it is well known that humans disagree with each other in making relevance judgments~\cite{Lesk_Salton_1968,Voorhees_SIGIR1998,Harman_2011,Cormack_etal_IPM2000}.
Due to budget and time limitations, we were not able to have the same topics annotated by multiple human assessors independently.
Without this, we are missing an important point of reference, as the divergence between human and LLM-assisted processes needs to be compared to human--human inter-annotator agreement.
Nevertheless, with respect to RQ3, the following conclusion seems warranted:

\begin{itemize}

\item[{\bf RQ3}] Human assessors appear to apply stricter relevance criteria than the current implementation of \umb.

\end{itemize}

\noindent That is, \umb often finds a passage to be more relevant than the NIST assessor.
This is in part because the LLM draws inferences that a human would consider unwarranted.
The opposite case is observed as well, but appears to be less common.
To be precise, this is a statement about the current state of LLMs with prompts used today (specifically, \umb).
Previous studies (e.g.,~\cite{Venanzi_etal_WWW2014}) have shown that human assessors may exhibit systematic tendencies to be more lenient or strict, and thus it is difficult to draw conclusions about humans and LLMs {\it in general}.
For example, we can perhaps alter LLM behavior with better prompting.

\section{Conclusions}

Recent studies have highlighted the potential of using LLMs for automating relevance assessments, as part of the broader ``LLM-as-a-judge'' literature~\cite{Bai:2212.08073:2022,Gao:2304.02554:2023,Cohen:2305.13281:2023,ZhengLianmin:2306.05685:2023}.
Production deployment of LLM-based assessments at Bing~\cite{ThomasPaul_etal_SIGIR2024} affirms this potential, but there remain many unknowns and different perspectives from the community~\cite{Faggioli_etal_ICTIR2023,Soboroff:2409.15133:2024}.
In this study, we have taken an important step in validating three approaches to using LLMs for relevance assessment, deployed {\it in situ} at a large-scale TREC evaluation.
Our results provide an empirically grounded validation study for academic TREC-style evaluations, serving as a foundation on which other researchers can build.
This study clarifies a few unknowns, but many questions remain unaddressed.

\section*{Acknowledgments}

This research was supported in part by the Natural Sciences and Engineering Research Council (NSERC) of Canada.
Additional funding is provided by Snowflake and Microsoft via the Accelerating Foundation Models Research program.

\bibliographystyle{ACM-Reference-Format}
\bibliography{rag24-qrels}

\balance

\end{document}